\documentclass[journal=jacsat,manuscript=article]{achemso}

\usepackage[version=3]{mhchem} 
\usepackage{gensymb}
\usepackage{textgreek}

\author{Daniel L. Druffel}
\author{Matthew G. Lanetti}
\author{Jack D. Sundberg}
\author{Jacob T. Pawlik}
\author{Madeline S. Stark}
\affiliation[First Affiliation]{Department of Chemistry, University of North Carolina at Chapel Hill, Chapel Hill, North Carolina 27599, United States}
\author{Carrie L. Donley}
\affiliation[Third Affiliation]{Chapel Hill Analytical and Nanofabrication Laboratory (CHANL), University of North Carolina at Chapel Hill, Chapel Hill, North Carolina, 27599, United States}
\author{Lauren M. McRae}
\author{Katie M. Scott}
\author{Scott C. Warren}
\affiliation[First Affiliation]{Department of Chemistry, University of North Carolina at Chapel Hill, Chapel Hill, North Carolina 27599, United States}
\alsoaffiliation[Second Affiliation]{Department of Applied Physical Sciences, University of North Carolina at Chapel Hill, Chapel Hill, North Carolina 27599, United States}
\email{sw@unc.edu}

\title[An \textsf{achemso} demo]
  {Synthesis and Electronic Structure of a Crystalline Stack of MXene Sheets}

\abbreviations{IR,NMR,UV}
\keywords{American Chemical Society, \LaTeX}

\begin{document}
\begin{abstract}

Despite the interest in MXenes in the last decade, all of the MXenes reported have a random mixture of surface terminations (-O, -OH, -F). In addition, restacked 3D films have turbostratic disorder and often contain ions, solvent, and other species in between their layers. Here we report Y$_2$CF$_2$, a layered crystal with a unit cell isostructural to a MXene, in which layers are capped only by fluoride anions. We directly synthesize the 3D crystal through a high-temperature solid-state reaction, which affords the 3D crystal in high yield and purity and ensures that only fluoride ions terminate the layers. We characterize the crystal structure and electronic properties using a combination of experimental and computational techniques. We find that relatively strong electrostatic interactions bind the layers together into a 3D crystal and that the lack of orbital overlap between layers gives rise to a description of Y$_2$CF$_2$ as slabs of MXene-like sheets electrically insulated from one another. Therefore, we consider Y$_2$CF$_2$ as a pure 3D crystalline stack of MXene-like sheets. In addition, Y$_2$CF$_2$ is the first transition metal carbide fluoride experimentally synthesized. We hope this work inspires further exploration of transition metal carbide fluorides, which are potentially a large and useful class of compositions. 

\end{abstract}

\section{Introduction}

Research into layered and 2D metal carbides such as MXenes has rapidly grown~\cite{Naguib2014} due to their remarkable properties and applications in electronic,~\cite{Zhao_M2015, Li_R2017} sensing,~\cite{Kim_S2018,Yu_X2015} and energy storage devices.~\cite{Naguib2012,Liang_X2015,Wang_X2015} Despite the interest in MXenes in the last decade, all of the MXenes reported have a random mixture of surface terminations (-O, -OH, -F), which complicates their study and use. The presence of multiple surface terminations on MXene sheets owes to their synthesis, which involves etching a MAX (M = metal; A = Al, Ga, Si, In; and X = C, N) phase with aqueous HF.~\cite{Naguib2011} Further complicating their study, restacked 3D films of MXenes have turbostratic disorder\cite{Ghidiu2017,Ghidiu2018} and often contain ions, solvent, and other species in between their layers.~\cite{Ren_C2015,Xia_Y2018}

Given the general interest in MXenes, it is remarkable that no pure (i.e. no -O, -OH), transition metal carbide fluoride has been reported. Of the transition metals, only Y,~\cite{Hwu_S2985,henn1996bulk,mattausch1995polytypism,ahn2005influence,kauzlarich1988two,kauzlarich1990y10i13c2,schaloske2009se,mattausch1994crystal,hinz1995crystal,simon1996supraleitung} Sc,~\cite{Hwu_S2985,jongen2006molecular,dudis1986synthesis,hwu1986metal,dudis1987two,hwu1985interstitial} Zr,~\cite{Hwu_S2985,smith1986four,smith1985stabilization} and W~\cite{strobele2010new} have been studied as pure metal carbide halides, and none as fluorides, leaving a large void in our knowledge of ternary phase diagrams. Our efforts in this work are to synthesize the first transition metal carbide fluoride, and to highlight a largely unexplored class of compositions that may find utility in optoelectronics.

Here we report Y$_2$CF$_2$, a layered crystal with a unit cell isostructural to a 2D MXene, in which layers are capped only by fluoride anions. Relatively strong electrostatic interactions bind the layers together into a 3D crystal. Therefore, we consider Y$_2$CF$_2$ as a pure 3D crystalline stack of MXene-like sheets. Instead of the etching approach, we describe an alternative synthetic method. We directly synthesize the 3D crystal through a high-temperature solid-state reaction in the absence of air and water. The synthesis affords the 3D crystal in high yield and purity and ensures that only fluoride ions terminate the layers. This allows us to identify several useful properties of Y$_2$CF$_2$: an indirect band gap at 1.6 eV, a direct band gap at 1.9 eV, a small ionization potential of 3.8 eV, and modest charge transport properties, with an electron effective mass of 2.0 and a hole effective mass of 4.0.  Perhaps most usefully, the realization of 3D crystalline MXenes will enable a comparison to their 2D MXene counterparts to better understand the influence of disorder and surfaces on their optoelectronic properties.

\section{Experimental section}

\subsection{The synthesis of Y$_2$CF$_2$}

Y$_2$CF$_2$ was synthesized by the solid state reaction of YF$_3$ (Sigma Aldrich, granules, 0.09\% oxygen, $<$0.02\% Cd, Co, Cr, Fe, Hg, Mn, Pb each) with Y metal (Alfa Aesar, $\sim$40 mesh powder, 99.6\% (REO)) and graphite (Sigma Aldrich, $\sim$100 mesh flakes, 99.9\%). The reagents were ground into a very fine powder in a stoichiometric ratio (total mass of a typical batch: 0.5--1.0 g) and pressed into a pellet under $\sim$0.56 GPa of pressure using a hydraulic press. The pellet was placed into an ampoule made of tantalum (Ta) metal welded closed at one side. A cap, also made of Ta, was hammered into the ampoule and sealed in argon by welding the cap and ampoule together with an electric arc. The Ta tubing, 99.95\% Ta seamless tubing, was purchased from Eagle Alloys in a 1 cm diameter and cut to 5 cm lengths. The caps were made of the same Ta. After welding the reagents inside a Ta ampoule, the Ta ampoule was sealed in a fused quartz ampoule under 1 mbar Ar. Then the ampoule was placed in a Lindberg Blue M tube furnace to 1513 K at a ramp rate of 10 K/min. The temperature was held at 1513 K for 3 days, then cooled to 1273 K over 10 hrs, then cooled to 1073 K over 5 hrs, and cooled to room temperature over 5 hours. The ampoule was then brought into the glovebox and opened with a pipe cutter. The pellet remained intact through the reaction. Vibrant green crystals grew on the surface of the pellet. The pellet revealed homogenous green crystals throughout when broken apart. For some experiments, samples were ball-milled for 15 minutes to reduce particle size. All materials were stored in a glovebox with an argon atmosphere (oxygen $<$ 0.01 ppm) and all synthetic steps were carried out under an argon atmosphere. 

\subsection{Material characterization}

Samples were maintained under an inert atmosphere during transport and loading procedures, unless otherwise stated. Powdered samples of Y$_2$CF$_2$ were characterized by X-ray diffraction at room temperature in a capillary transmission geometry across a range of 20-150$\degree$ using Cu K$\alpha$ ($\lambda$ = 1.54056 {\AA}) in a Rigaku Smartlab diffractometer. A reference scan was run with silicon (\textit{a} = 5.430 {\AA}) as an internal standard. The peaks were not corrected as the height of the sample was already correct. The patterns were indexed using Treor~\cite{werner1985treor} implemented in the Match! software.~\cite{Match!} The patterns were refined using the PDXL software from Rigaku.~\cite{version20112} Thin films were characterized by XPS using a Kratos Axis Ultra Delay-Line Detector (DLD) spectrometer with monochromatic Al K$\alpha$ source. For these measurements, the powder was pressed into indium substrates. High resolution XPS data was collected at a pass energy of 20 eV, and a charge neutralizer was used for charge compensation.  All data was corrected to the C 1s peak at 284.6 eV. The samples were imaged using a Hitachi S-4700 cold cathode field emission SEM at an accelerating voltage of 2 kV with an Oxford EDS detector. An accelerating voltage of 20 kV was used for EDS measurements. Samples were loaded onto the SEM sample holders in the glovebox and transported to the instrument under inert atmosphere, however, the samples were exposed to air for a few seconds during the transfer into the air lock of the instrument. UV-Vis spectroscopy was performed in a transmission geometry through thin ($\sim$0.5mm-thick) discs made of mixtures of Y$_2$CF$_2$ and KBr. The Y$_2$CF$_2$ powder was diluted with spectroscopic grade KBr and ground very finely in a ball mill. The mixture was ground until homogenous, then pressed into discs using a hydraulic press. The background due to scattering from KBr was subtracted by pressing a number of films of KBr of varying thickness to estimate its effective attenuation as a function of thickness and subtracting a weighted fraction from the acquired spectra. All discs appeared red, not green. The spectra were collected using a Cary 5000 UV-Vis spectrometer with the DRA-2500 internal integrating sphere accessory. The reflection spectra of Y$_2$CF$_2$ particles from 400--850 nm were obtained using a Craic UV-vis-NIR Microspectrophotometer (MSP) with a polychromatic Xe source and an EC Epiplan Neofluar LD objective (NA: 0.55). The aperture size was 14.7$\times$14.7 microns except where noted. 

\subsection{Computational experiments}

Density functional theory (DFT) calculations were performed using the CASTEP~\cite{Clark2005} code with plane-wave basis set approximations. To describe core electrons, ultrasoft pseudopotentials~\cite{Vanderbilt1990} were used with a 600 eV cut-off energy. For this cut-off energy, calculations were convergent with dE\textsubscript{tot}/dlnE\textsubscript{cut} less than 0.01 meV/atom. Geometry optimizations of all structures were first performed. A GGA PBE-sol functional~\cite{perdew2008restoring} was used for the exchange-correlation contribution to total energy and Tkatchenko and Scheffle's correction~\cite{Tkatchenko2009} was used to account for long-range dispersion forces for all calculations except where noted. A Monkhorst-Pack grid~\cite{Monkhorst1976} of 16$\times$16$\times$8 $k$-points were used and structures were optimized using a BFGS algorithm with a convergence tolerance of 5.0 $\times$ 10$^{-7}$ eV/atom for energy, 0.01 eV/{\AA} for max force, 0.02 GPa for max stress, and 5.0 $\times$ 10$^{-4}$ {\AA} for max displacement. The ``the thin-slab approach"~\cite{fall1999deriving} was used to calculate the ionization potential of bulk Y$_2$CF$_2$. For the calculations of ionization potential, a denser $k$-point mesh (32$\times$32$\times$16) was used. To prevent self-interactions between periodic images, a 12 {\AA} vacuum space was included in the \textit{z}-direction. The band gap and density of states were calculated using the HSE06 functional~\cite{HSE06} with norm-conserving pseudopotentials,~\cite{Hamann1979} a 850 eV cut-off energy, and a $k$-point mesh of 32$\times$32$\times$16. The fixed-composition USPEX calculation was carried out via four serial geometry optimizations of each candidate structure. The geometry optimizations utilized the GGA PBE functional with increasing quality, where the fourth and final optimizations were converged to  5$\times$10$^{-5}$ eV/atom using an 440 eV cutoff energy. For accurate comparison of candidate stabilities, geometry optimization were followed by a single, higher-quality energy calculation converged to 5$\times$10$^{-6}$ eV/atom deemed sufficient to isolate reasonable Y$_2$CF$_2$ structure types. The search identified the minimum Y$_2$CF$_2$ structure from among roughly 400 candidates tested across 8 generations.

\section{Results and discussion}

\subsection{Synthesis and crystal structure}

We combined yttrium, carbon, and fluorine in a stoichiometry of Y$_2$CF$_2$ at 1240 $\degree$C in sealed tantalum ampoules according to the following reaction:

\begin{equation}
    4 \text{Y} + 3\text{C} + 2 \text{YF}_3 \rightarrow 3 \text{Y}_2\text{CF}_2
    \label{eq:Synth_Y2CF2_1}
\end{equation}

The product is a pellet of beautiful green crystals (Figure \ref{eq:Synth_Y2CF2_1}a) that easily grinds into a powder (Figure \ref{eq:Synth_Y2CF2_1}b). Unfortunately, the synthesized crystals were too small and inter-grown to perform single crystal X-ray diffraction measurements. Therefore, we combined two independent approaches to determine the crystal structure.

First, we performed powder X-ray diffraction experiments of the product and obtained a diffraction pattern (Figure \ref{eq:Synth_Y2CF2_1}c), which indexed as a trigonal crystal structure with space group P$\overline{3}$m1-D$_{3d}$. Then, we refined the pattern by the Rietveld method~\cite{rietveld2014rietveld} (Table \ref{tbl:Y2CF2_refinement}). The analysis confirms that the trigonal space group P$\overline{3}$m1 is a good fit, with lattice constants \textit{a}=3.6651 (3) {\AA} and \textit{c} = 6.29697 (7) {\AA} at room temperature (Figure \ref{eq:Synth_Y2CF2_1}c, Table \ref{tbl:Y2CF2_refinement}). 

\begin{figure}[h!]
    \centering
    \includegraphics[width=10.16cm]{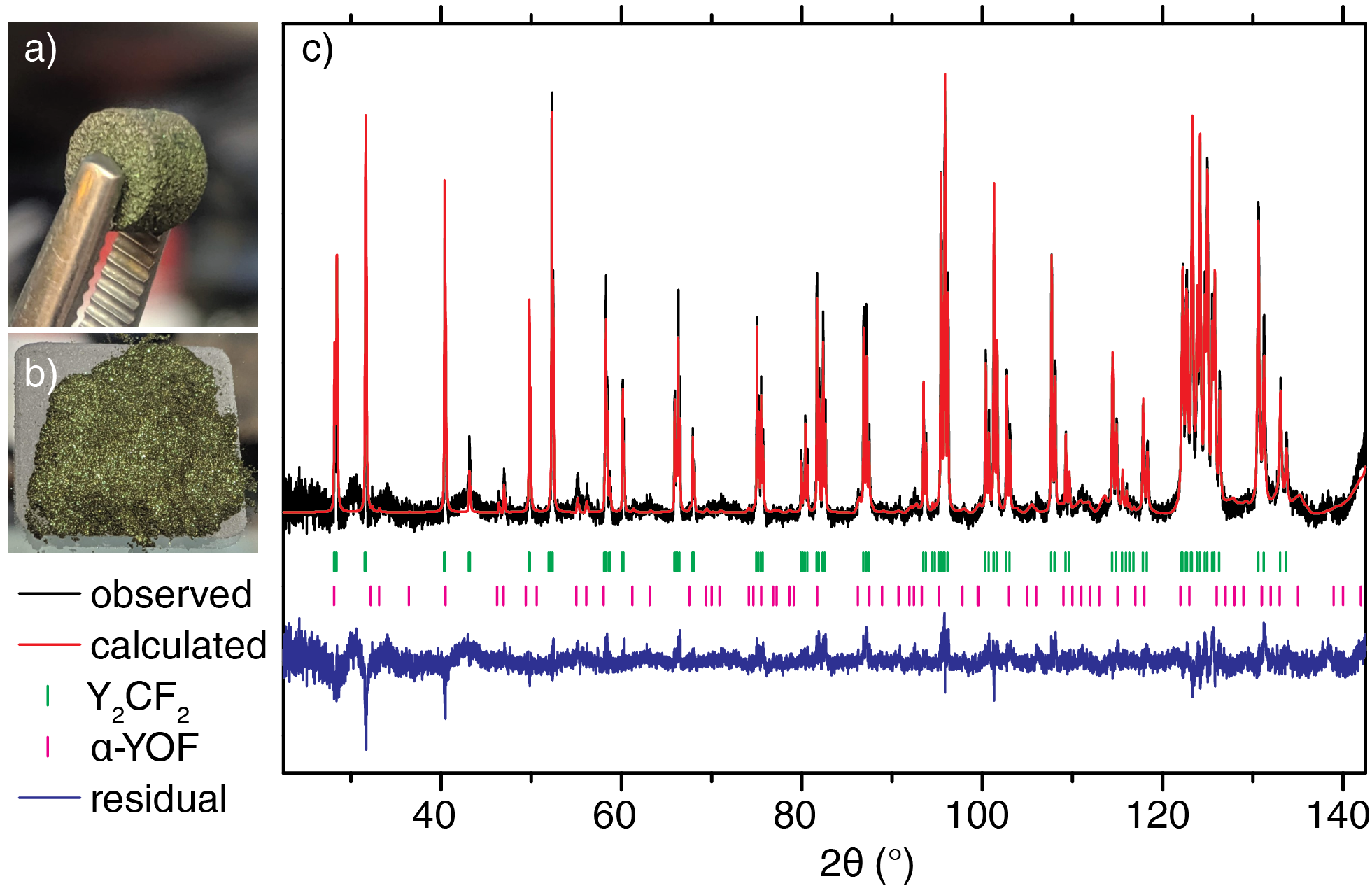}
    \caption[X-ray diffraction pattern for Y$_2$CF$_2$]{Observed, calculated and difference profiles for Y$_2$CF$_2$ at 300 K measured on a Rigaku Smartlab diffractometer with a Cu K$\alpha$ source, $\lambda$ = 1.5406 {\AA}. The pattern was refined to two phases, a Y$_2$CF$_2$ phase and a $\alpha$-YOF phase (Space group P4/nmm, \textit{a} = 3.922 (2) {\AA}, \textit{c} = 5.409 (4) {\AA}), which made up less than 4{\%} by mole of the sample. Vertical tick marks indicate calculated reflection positions. The broad maxima in the background are due to diffuse scattering from the quartz ampoule.}
    \label{fig:Y2CF2_XRD}
\end{figure}

\begin{figure}[b!]
    \centering
    \includegraphics[width=12.16cm]{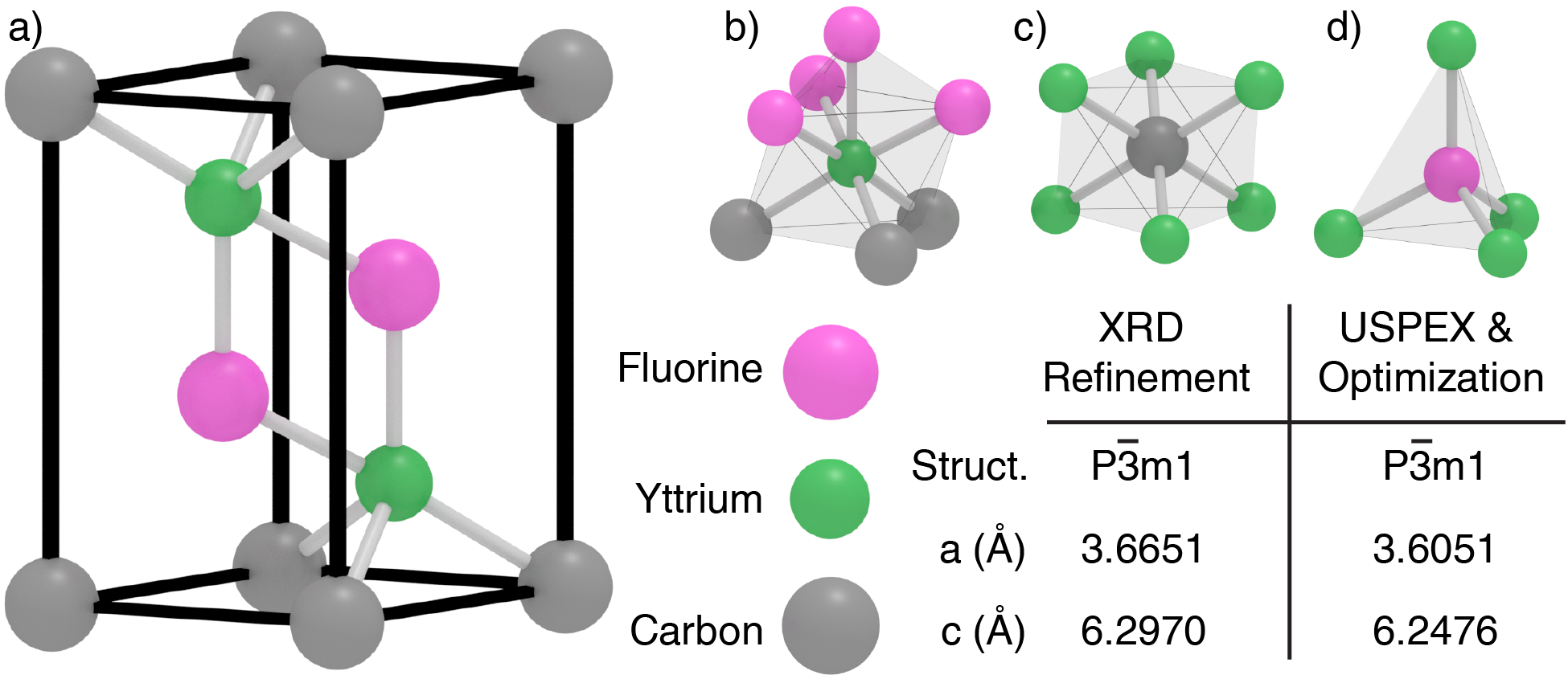}
    \caption[Crystal structure, bonding polyhedral, and lattice constants for Y$_2$CF$_2$]{Crystal structure, bonding polyhedral, and lattice constants for Y$_2$CF$_2$. The crystal structure and lattice constants obtained independently from two methods: 1) experimental synthesis and Rietveld refinement of X-ray diffraction data and 2) USPEX, a DFT-based search algorithm to find the most stable crystal structure for a given composition. After using USPEX to identify the structure, the structure was optimized using the PBEsol functional as described in the computational experimental section.}
    \label{fig:Y2CF2_Crystal}
\end{figure}

Independently, we used an evolutionary search algorithm, USPEX,~\cite{glass2006uspex} to explore the stability of crystal structures for the composition Y$_2$CF$_2$ computationally. The algorithm computes the enthalpy of hundreds of candidate crystal structures and uses similarities in bond distances, coordination, symmetry, and a number of other identifiers to find the most stable phase for a given composition. Using the algorithm, we found that the P$\overline{3}$m1 structure is the most enthalpically favorable phase for the composition Y$_2$CF$_2$, in agreement with our diffraction experiments. This structure is more stable than the second-most stable structure by 46 meV/atom. We then performed a geometry optimization calculation using the PBEsol~\cite{perdew2008restoring} functional to calculate the lattice constants. We calculated constants very close to the experimental value (1.6\% shorter in \textit{a} and 0.8\% shorter in \textit{c}, Figure \ref{fig:Y2CF2_Crystal}b).

\begin{table}[h]
  \small
  \centering
  \caption[Refinement of the structure of Y$_2$CF$_2$]{Final parameters, R factors and interionic distances for Y$_2$CF$_2$ at 300 K. Space group P$\overline{3}$m1. \textit{a} = 3.6651 (3) {\AA}, \textit{c} = 6.29697 (7) {\AA}. Volume = 73.2536 {\AA}$^3$, density = 4.394 g cm$^{-3}$ Rietveld analysis standard agreement indices are the residual factor Rp = 2.5\%, the weighted residual factor Rwp = 4.97\% and the goodness of fit factor S = 1.96. $U$, $V$, $W$ (deg$^2$): 0.000, -0.021, 0.000} \vspace{5px}
  \begin{tabular}[c]{ c c c c c c c c }
    \hline
    \rule{0pt}{3ex} Atom & Site symmetry & X & Y & Z & B ({\AA}$^2$) & N & Occupancy\\
    \hline
    \rule{0pt}{3ex} Y & 2d \: 3m & $\frac{1}{3}$ & $\frac{2}{3}$ & 0.21450 (10) & 0.526 (12) & 2 & 1.00 (1) \\
    C & 1a \: $\overline{3}$m & 0 & 0 & 0 & 0.1 (4) & 1 & 0.93 (3)\\
    F & 2d \: 3m & $\frac{1}{3}$ & $\frac{2}{3}$ & 0.0.6031 (5) & 0.555 (11) & 2 & 1.00 (6)\\
    \hline
  \end{tabular}
  \label{tbl:Y2CF2_refinement}
\end{table}

We also note that many yttrium compounds are isostructural to gadolinium and holmium compounds (e.g. YOCl~\cite{garcia1985low}, YCl$_3$~\cite{templeton1954crystal}, YF$_3$~\cite{zalkin1953crystal}) because of the similarity in the crystallographic ionic radius of Y (104 pm)~\cite{shannon1970revised} to that of Gd and Y (108 and 104 pm respectively).~\cite{shannon1970revised} In support of our indexing, we found known gadolinium and holmium compounds (Gd$_2$CF$_2$~\cite{mattausch1991erste} and Ho$_2$CF$_2$~\cite{cockcroft1992structure}) isostructural to Y$_2$CF$_2$ with very similar lattice constants.

The crystal structure of Y$_2$CF$_2$ (Figure \ref{fig:Y2CF2_Crystal}a) is similar to that of MXenes. Six Y cations octahedrally coordinate each carbon atom and the octahedra (Figure \ref{fig:Y2CF2_Crystal}c) form into a two-dimensional layer, just like the 2D Ti$_2$C(-O, -OH, -F) MXene.~\cite{come2012non} However, in the Y$_2$CF$_2$ structure, each Y cation also coordinates to four F$^-$, forming face-capped octahedra (Figure \ref{fig:Y2CF2_Crystal}b). Thus, in Y$_2$CF$_2$, double layers of fluoride ions bind adjacent metal-carbide layers forming a crystalline 3D stack of MXene-like sheets.

\begin{figure}[h]
    \centering
    \includegraphics[width=7.62cm]{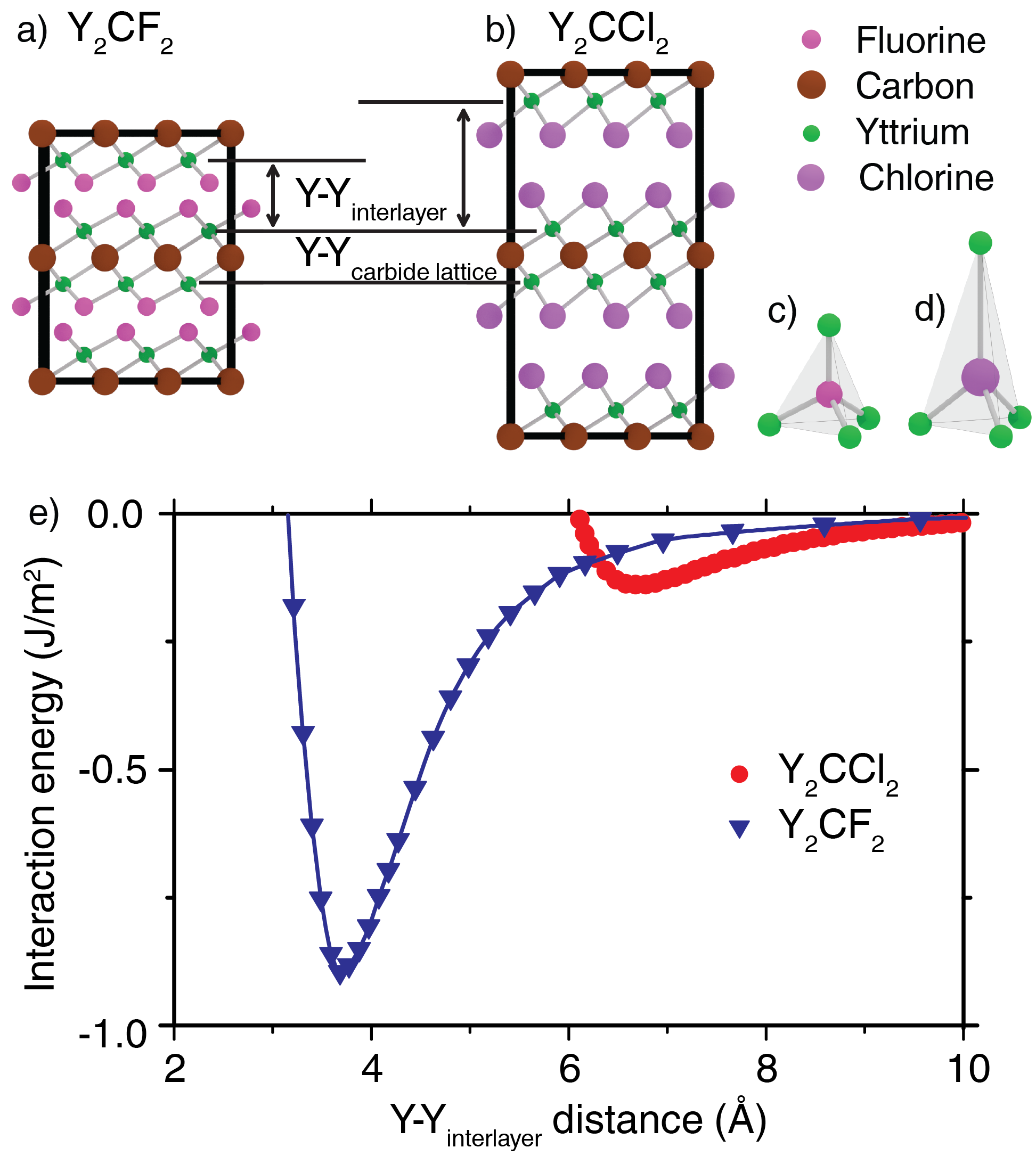}
    \caption[Comparison of the crystal structure of Y$_2$CF$_2$ to Y$_2$CCl$_2$]{Comparison of the crystal structure of Y$_2$CF$_2$ to Y$_2$CCl$_2$ showing a) the unit cells along the [100] axis, b) the tetrahedral bonding polyhedron for yttrium and fluorine in Y$_2$CF$_2$ and the same polyhedron drawn for Y$_2$CCl$_2$ with the elongated Y--Cl distance in the \textit{z}-direction, and c) the interlayer binding energy for Y$_2$CF$_2$ (blue) and Y$_2$CCl$_2$ (red) calculated using DFT.}
    \label{fig:Y2CF2_Y2CCl2}
\end{figure}

Naturally, we became interested in the interlayer bonding of Y$_2$CF$_2$, which is distinct from the 2D nature of MXenes. The F coordinate to 4 Y with four nearly symmetric distances (Figure \ref{fig:Y2CF2_Y2CCl2}c), three to the Y in the adjacent layer (2.408 {\AA}) and one to the Y in the opposite layer at an only slightly elongated distance (2.447 {\AA})(Table \ref{tbl:Y2CF2_atomic_dist}). This suggests that the interlayer bonding is relatively strong (as strong as the Y--F bonds) and highly ionic in nature.

\begin{table}[t]
  \small
  \centering
  \caption[Interatomic distance in the crystal structure of Y$_2$CF$_2$]{Interatomic distances in the crystal structure of Y$_2$CF$_2$, determined by the refinement of powder X-ray diffraction data. All distances are reported in {\AA}.} \vspace{5px}
  \begin{tabular}[c]{ c l }
    \hline
    \rule{0pt}{3ex} Atoms & Interatomic distances {\AA} \\
    \hline
    \rule{0pt}{3ex} Y--Y & 3.4315 (2), 3.6651 (3), 4.1720 (1) \\
    Y--F & 2.4077 (3), 2.4470 (1) \\
    F--F & 2.4827 (7) \\
    Y--C & 2.5104 (6) \\
    F--C & 3.2747 (6) \\
    \hline
  \end{tabular}
  \label{tbl:Y2CF2_atomic_dist}
\end{table}

This electrostatic interlayer bonding contrasts with that of other known layered metal carbide halides. For example, in Y$_2$CCl$_2$,~\cite{Hwu_S2985} which shares the same crystal structure (Figure \ref{fig:Y2CF2_Y2CCl2}b, Cl coordinates only to 3 Y, all in the adjacent layer (2.755 {\AA}). The distance to the Y in the opposite layer is nearly double (4.770 {\AA}), far too long for any Y--Cl bonding interaction (Figure \ref{fig:Y2CF2_Y2CCl2}d). Instead, the primary interlayer interactions in Y$_2$CCl$_2$ are weak van der Waals forces.

Using DFT and the PBEsol functional we calculated the binding energy---the energy difference between the equilibrium bulk separation and the state in which the planes are separated by an infinite distance, as approximated by a 15 {\AA}-thick vacuum-gap (Figure \ref{fig:Y2CF2_Y2CCl2}e). The binding energy for Y$_2$CF$_2$ is 0.90 J/m$^2$, over 6$\times$ that of Y$_2$CCl$_2$ (0.14 J/m$^2$) and 3$\times$ that of graphite (0.34 J/m$^2$).~\cite{Spanu2009} Electrostatic interactions between the F anion and Y cation likely account for the greater binding energy. This leads us to conclude that Y$_2$CF$_2$ is an ionic crystal in the out-of-plane direction. Still, the binding energy of Y$_2$CF$_2$ is less than that of other layered crystals, for example Ca$_2$N (1.11 J/m$^2$), that have been exfoliated successfully into 2D flakes.~\cite{Druffel2016} These calculations suggest that it may be possible to exfoliate Y$_2$CF$_2$ into 2D flakes, although with somewhat greater difficulty than van der Waals crystals like Y$_2$CCl$_2$ or graphite.

\subsection{Composition and purity}

We sought to confirm the composition of our sample, as some metal carbides can sustain a variety of non-stoichiometric compositions. We used energy-dispersive X-ray spectroscopy (EDS) to quantify the amount of yttrium and fluorine (Figure S2) and measured a ratio of 1:0.98 $\pm$ 0.05 Y:F in agreement with the stoichiometry of Y$_2$CF$_2$. These results also agree with the refinement from powder X-ray measurements, which yielded an occupancy of 1.00 $\pm$ 0.01 and 1.00 $\pm$ 0.06 for yttrium and fluorine respectively in each of their 2d sites (Table \ref{tbl:Y2CF2_refinement}). Unfortunately, our EDS instrument does not quantify carbon reliably because the signal is obscured by electronic noise. Therefore, we quantified carbon by refining the occupancy of the 1a (carbon) site during the Rietveld refinement, which revealed that only 93 $\pm$ 3 \% of carbon sites are occupied. We conclude that the composition Y$_2$CF$_2$ represents the synthesized phase, though the real structure has carbon vacancies, Y$_2$C$_{0.93}$F$_2$.

\subsection{Electronic structure of Y$_2$CF$_2$}

To investigate the effect of interlayer ionic bonding on the electronic structure of stacked MXene sheets, we examined the material's optoelectronic properties using both experiment and computation. For all of the DFT calculations of the electronic structure, we used the HSE06 hybrid functional,~\cite{HSE06} which we selected because of its suitability for estimating band gaps.~\cite{}

\begin{figure}[h]
    \centering
    \includegraphics[width=16.5cm]{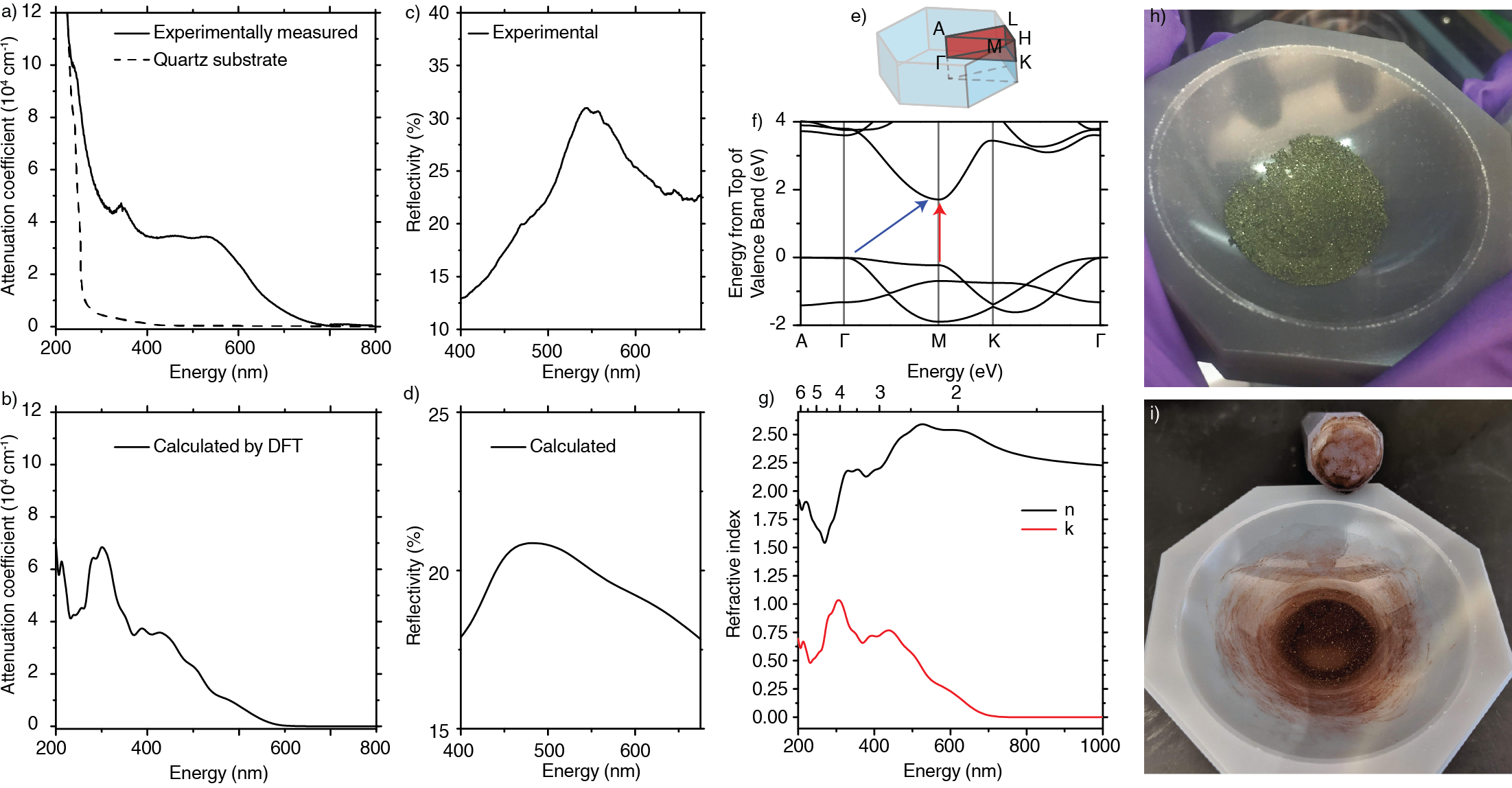}
    \caption[Optical properties of Y$_2$CF$_2$.]{Optical properties of Y$_2$CF$_2$. The attenuation coefficient a) measured experimentally as described in the experimental as the slope of the attenuation as a function of effective thickness, which inherently subtracts the contribution from quartz substrate; however, at wavelengths shorter than 250 nm the signal was dominated and obscured by that of the quartz. The attenuation coefficient b) calculated by DFT, the reflectivity c) measured experimentally and d) calculated by DFT. e) The first Brillouin zone for Y$_2$CF$_2$ with the irreducible Brillouin zone highlighted in red and f) the electronic band structure where the M--$\Gamma$ vector runs along (010), while the $\Gamma$--A vector runs along (001). g) The optical constants, n and k, calculated by DFT. Photograph of h) green Y$_2$CF$_2$ before grinding and i) red Y$_2$CF$_2$ after grinding.}
    \label{fig:Y2CF2_Optics}
\end{figure}

First, we performed transmission measurements to measure the absorption coefficient (Figure S4a). We observe a gradual increase in absorption beginning at $\sim$775 nm (1.6 eV), which we attribute to the band gap. The absorption increases and plateaus from 400-550 nm at a magnitude of 3.4 $\times$ 10$^4$ cm$^{-1}$. The absorption coefficient peaks at 340 nm and again at 240 nm, though the peak at 240 is obscured by the onset of attenuation of our quartz substrate. As shown in figure \ref{fig:Y2CF2_Optics}b, our model calculates a gradual increase in absorption beginning at 700 nm (1.7 eV), in excellent agreement with our experimental measurements. The absorption coefficient plateaus from 375-450 nm at a magnitude of 3.4 $\times$ 10$^4$ cm$^{-1}$, nearly identical to our experiment. Finally, the model predicts peaks at 300 nm and 210 nm, redshifted only 40 nm from our experimental data.

Next, we performed specular-reflectance measurements on single facets of individual Y$_2$CF$_2$ crystals (Figure S4) using a microspectrophotometer. To obtain a representative spectrum, we averaged the spectra from individual crystallites together (Figure \ref{fig:Y2CF2_Optics}c). The representative spectrum has a peak at 540 nm, corresponding to yellow-green light. To compare with our model of the electronic structure, we calculated the reflectivity for the bulk Y$_2$CF$_2$ (Figure \ref{fig:Y2CF2_Optics}d), which shows a single broad peak at 500 nm, again redshifted only 40 nm from experiment. We do note that the intensity of the reflection in the experiment is roughly 50\% larger than that in the computation. While the calculation provides qualitative information, the calculation does not consider all of the complexity of the real system---chemical differences at the surface, an interface with a changing dielectric, or the convolution of reflection and absorption events. Therefore, we attribute the difference in magnitude to these factors. Despite this difference, the model reproduces the major feature of the experiment very well.

We calculated the electronic band structure for the first Brillouin zone of Y$_2$CF$_2$ shown in Figure \ref{fig:Y2CF2_Optics}e,f. In agreement with our experimental measurements, we calculate that Y$_2$CF$_2$ is a semiconductor with a 1.7 eV indirect gap (from $\Gamma$--M) and a 1.9 eV direct gap (M--M)(Figure \ref{fig:Y2CF2_Optics}f). From the band structure, we can understand that the low absorption coefficient at 700 nm (Figure \ref{fig:Y2CF2_Optics}a,b) is due to the indirect nature of the transitions and the low density of states in the conduction band. Moreover, we gain information about the band dispersion. The bottom of the conduction band has a parabolic band dispersion, which should correspond to a modest effective mass (2.0) for electrons in the conduction band. In contrast, the valence band is relatively flat giving rise to a higher hole effective mass (4.0). Interestingly, the band has almost no dispersion along the path $\Gamma$--A (Figure \ref{fig:Y2CF2_Optics}f) indicating very limited overlap of orbitals in the \textit{z}-direction. This is likely due to the strong localization of electrons around the nucleus of the fluorine atoms, which separate layers of yttrium and carbon atoms.~\cite{albright2013orbital} Therefore, the ionic bonding between MXene sheets effectively insulates the sheets from one another electronically, which will give rise to highly anisotropic optoelectronic properties.

Because of the excellent agreement between theory and experiment, we present the optical constants (Figure \ref{fig:Y2CF2_Optics}g), $n$ and $k$, calculated by DFT using the HSE06 functional with confidence that the energy-dependence of the optical constants and that the magnitude of $k$ represent the synthesized Y$_2$CF$_2$. Due to the difference in magnitude between experiment and theory we are less confident in the magnitude of $n$.

This electronic structure also explains an observation about the perceived color of Y$_2$CF$_2$. When we ground the green crystals (Figure \ref{fig:Y2CF2_Optics}g) in a glovebox in the absence of air or water, the material transforms into a red powder shown in Figure \ref{fig:Y2CF2_Optics}i. By examining the electronic structure, we conclude that the initially perceived green color is due to specular reflection of green light from flat crystalline surfaces, as evidenced by the peak in reflection at 540 nm in Figure \ref{fig:Y2CF2_Optics}c. By grinding the crystals into small particles, we destroy the flat surfaces and the specular reflection of green light. Once ground, absorption, originating from the band gap at 775 nm, dominates the apparent color of the material. To test this hypothesis, we ball milled the crystals to a fine powder with a particle size qualitatively determined as less than 5 \textmu m (Figure S6). We confirmed, via pXRD, that the crystal structure is preserved after ball milling (Figure S7). Then we measured the reflectance of the finely ground sample using a microspectrophotometer (Figure S8). We found that the peak at 540 nm, observed for the green crystals, was not present in the ball-milled sample.

\subsection{Sensitivity to oxygen}

This understanding of Y$_2$CF$_2$'s electronic structure also led us to understand its sensitivity to oxygen and water. We removed samples of crystalline, green Y$_2$CF$_2$ from the glovebox, exposing the crystals to air and water for one month. After two weeks, the once-green crystals appeared red (Figure S9). Examining the crystals in the microspectrophotometer in reflectance mode, we found that the peak in reflectance at 540 nm corresponding to the green color was not present. Instead, we observe rainbow-striping (Figure S11) and periodic oscillating signals in the reflectance spectra (Figure S11), which are indicative of thin-film optical interference. The formation of a thin-film on the surface of many semiconductors is commonly caused by a slow reaction with air or water, which forms a thin oxide coating.~\cite{Kuntz2017,spitzer1973electrical,Morita_1994} Therefore, we suspected that Y$_2$CF$_2$ in air develops a thin film of an oxide or related species.

\begin{figure}[h!]
    \centering
    \includegraphics[width=16.5cm]{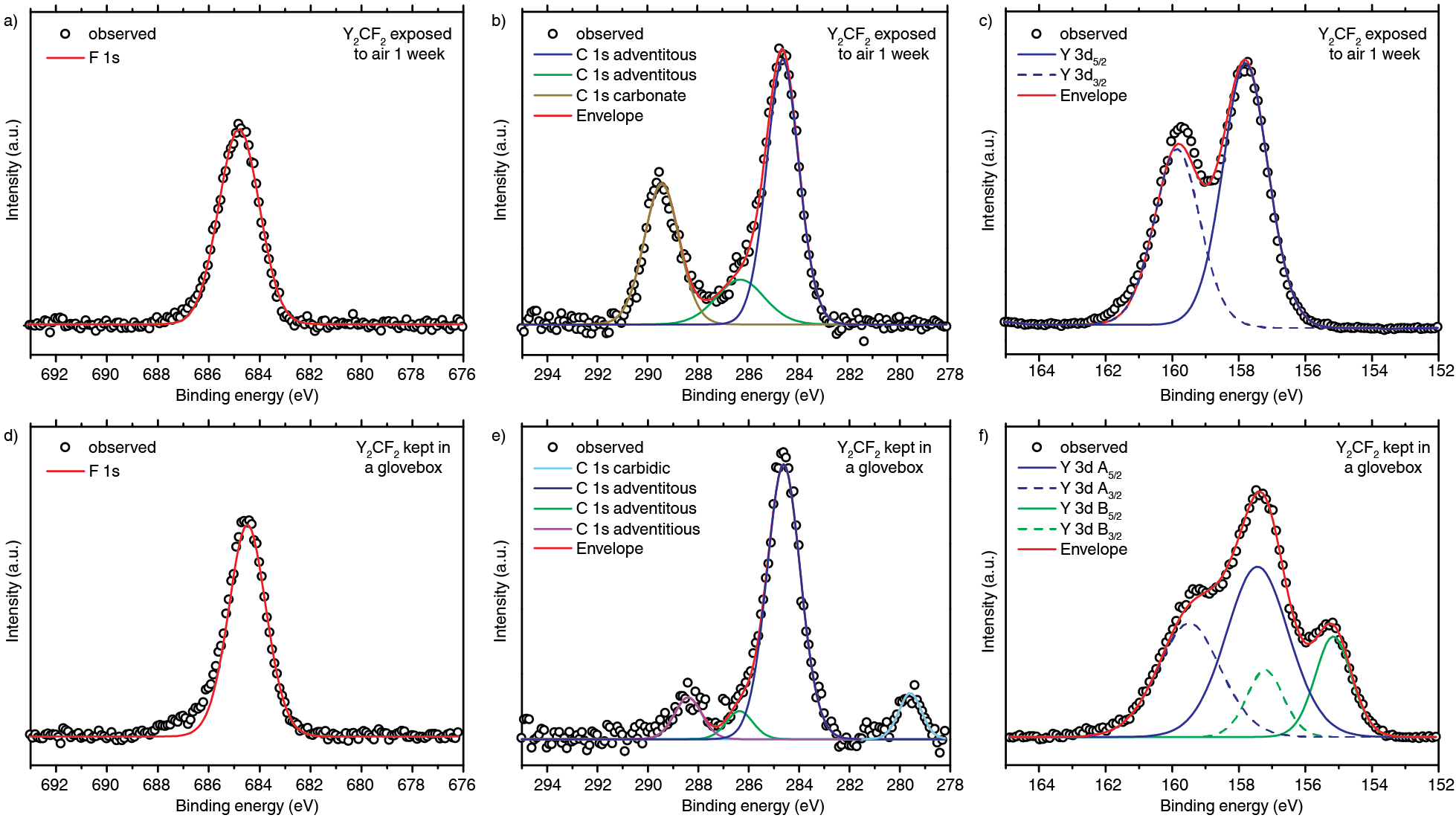}
    \caption[X-ray photoemission spectroscopy experiments of Y$_2$CF$_2$.]{X-ray photoemission spectra from Y$_2$CF$_2$ exposed to air for 1 week for the a) fluorine \textit{1s}, b) carbon \textit{1s}, c) yttrium \textit{3d} core electron binding energies. d) The spectrum for carbon \textit{1s} core electrons from air-exposed Y$_2$CF$_2$ after sputtering. Spectra from Y$_2$CF$_2$ kept in a glovebox for the e) yttrium \textit{3d} and carbon \textit{1s} core electron binding energies.}
    \label{fig:Y2CF2_XPS}
\end{figure}

To test whether the Y$_2$CF$_2$ crystals react with oxygen or water, we performed X-ray photoemission spectroscopy (XPS) on crystals exposed to air for 1 week and on crystals kept in the glovebox. We examined the yttrium \textit{3d}, carbon \textit{1s}, fluorine \textit{1s}, and carbon \textit{1s} core electron binding energies (Table \ref{tbl:Y2CF2_XPS}). For the sample exposed to air for 1 week, we observed a single peak at a binding energy of 684.8 eV (Figure \ref{fig:Y2CF2_XPS}a) corresponding to the F \textit{1s} core binding energy, in agreement with the binding energy of fluoride.~\cite{PhysRevB.12.5872} However, the Y:F ratio measured through XPS quantitative analysis is 2.0:1, suggesting a deficiency of fluorine at the surface. In the Y \textit{3d} spectrum we observe a doublet with a binding energy of 157.7 eV (Figure \ref{fig:Y2CF2_XPS}c). The position of this peak matches that of yttrium carbonate (Y$_2$C$_3$O$_9$)\cite{Vasquez1990, Dubbe1997} and is too high to be the expected Y$^{2+}$ oxidation state ($\sim$156 eV).~\cite{Fujimori1984,MONGSTAD2014270} In addition, the spectrum of the carbon \textit{1s} core (Figure \ref{fig:Y2CF2_XPS}b) reveals two distinct peaks, one centered at 284.6 eV, which is adventitious carbon and one centered at 289.4 eV, which matches that of a carbonate(Figure \ref{fig:Y2CF2_XPS}b).~\cite{doi:10.1063/1.119679,Dubbe1997} Moreover, through quantitative XPS analysis, we measure the ratio of Y:O to be 1.0:2.1, which can only be explained by a carbonate, not Y$_2$O$_3$ or YOF. Therefore, the data suggest the surface of the sample exposed to air has oxidized, at least partially, to Y$_2$C$_3$O$_9$.

\begin{table}[t]
  \small
  \centering
  \caption[X-ray photoemission spectra from samples of Y$_2$CF$_2$ kept in different environments]{X-ray photoemission spectra from Y$_2$CF$_2$ exposed to air for 1 week, after sputtering, and from Y$_2$CF$_2$ kept in an Ar-filled glovebox (oxygen $<$ 0.01 ppm) for 1 week. All binding energies and full-width half-max values given in eV. $^a$All binding energies were referenced to adventitious carbon (284.6 eV).} \vspace{5px}
  \begin{tabular}[c]{ c | l c l c c }
    \hline
    \rule{0pt}{3ex} Sample & Atom & Core & Binding Energy$^a$ & FWHM & Atomic \% \\
    \hline
    \rule{0pt}{3ex} Air-exposed & Y & 3d$_\frac{5}{2}$ & 157.6 & 1.56 & 21.3\\
    & C$_\text{adv}$ & 1s & 284.6$^a$, 286.3 & 1.33, 2.24 & 17.1 \\
    & C$_\text{carbonate}$ & 1s & 289.4 & 1.63 & 7.9 \\
    & F & 1s & 684.8 & 1.81 & 10.5 \\
    & O & 1s & 531.2 & 1.88 & 43.2 \\
    \hline
    \rule{0pt}{3ex} Air-exposed & Y & 3d$_\frac{5}{2}$ & 157.6 & 4.06 & 33.8\\
    sputtered & C$_\text{adv}$ & 1s & 284.6$^a$, 286.0 & 1.67, 1.76 & 5.3 \\
    & C$_\text{carbonate}$ & 1s & 289.9 & 1.57 & 1.9 \\
    & F & 1s & 685.0 & 2.11 & 16.9 \\
    & O & 1s & 531.3 & 3.91 & 42.1 \\
    \hline
    \rule{0pt}{3ex} Glovebox & Y & 3d$_\frac{5}{2}$ & 155.3 & 1.27 & 6.7\\
    & Y & 3d$_\frac{5}{2}$ & 157.6 & 2.19 & 19.4\\
    & C$_\text{carbide}$ & 1s & 279.6 & 1.04 & 4.5\\
    & C$_\text{adv}$ & 1s & 284.6$^a$, 286.3 & 1.25, 1.55 & 27.4 \\
    & F & 1s & 684.4 & 1.75 & 18.8 \\
    & O & 1s & 531.2 & 2.33 & 27.1 \\
    \hline
  \end{tabular}
  \label{tbl:Y2CF2_XPS}
\end{table}

We attempted to clean the surface of the air-exposed sample by argon ion sputtering. Both the oxygen \textit{1s} and carbon \textit{1s} (Figure \ref{fig:Y2CF2_XPS}d) signals decrease in intensity relative to the yttrium. The Y:O ratio decreases to 1:1.2. This suggests that the oxidation is a surface effect. For additional support of that hypothesis, we performed XRD measurements on the samples exposed to air (Figure S10). We found the crystal structure remained intact after exposure to air for 1--4 weeks, which supports the conclusion that the oxidation affects the surface and not the bulk of the sample.

For the sample kept in a glovebox, the sample was briefly exposed to air during the transfer into the XPS instrument. We attempted this measurement multiple times and found that the sample is prone to charging. Here, we present the data from the experiment that showed the least charging. The F \textit{1s} spectrum (Figure \ref{fig:Y2CF2_XPS}d) has a single peak at 684.4 eV similar to that of the air-exposed sample (Table \ref{tbl:Y2CF2_XPS}). We attribute the small tail at higher binding energies in the F \textit{1s} spectrum to this charging. The carbon \textit{1s} spectrum (Figure \ref{fig:Y2CF2_XPS}e) shows one main peak, which we attribute to adventitious carbon (Table \ref{tbl:Y2CF2_XPS}) and which we used to reference the data. Referencing to any other carbon peak yields unreasonable values for the other peaks in other spectra. In addition to the adventitious carbon, we observe a peak at a very low binding energy (279.6 eV). Carbidic carbon in similar materials, for example, Sc$_2$CCl$_2$ (281.8),~\cite{Hwu_S2985} Ti$_2$C (281.5), Zr$_2$C (281.1), and HfC (280.8), gives rise to peaks at low binding energies, but none quite as low as what we observe here. Therefore, we considered other possibilities that the signal could be due to other elements, which have signals at similar binding energies (Os \textit{4d}, Ru \textit{3d}, Sr \textit{3p}, Tb \textit{4p}, and Cl \textit{2s}). We rule out Os, Sr, Tb, and Cl becuase the XPS survey scan reveals that the most intense peaks for these elements are absent (Os \textit{4f} at 50--55 eV, Sr \textit{3d} and \textit{3s} at 130--135 eV and 357 eV respectively, Tb \textit{4d} at 150 eV, and Cl \textit{2p} at 208 eV). The most intense peak for Ru is the \textit{3d} peak. Therefore, we cannot rule out Ru directly. However, we have no reason to suspect the incorporation of Ru into our sample and its weaker peaks are absent (Ru \textit{4s} at 75 eV, Ru \textit{3p} at 463 eV, and Ru \textit{3s} at 586 eV). Therefore, we assign the peak at 279.6 eV as carbidic carbon in Y$_2$CF$_2$, consistent with a highly reduced form of carbon. We also note that we do not observe the signal at higher binding energies of a carbonate species for the sample kept in the glovebox. Finally, in the Y \textit{3d} spectrum, we observe two distinct yttrium signals (Figure \ref{fig:Y2CF2_XPS}f). One contribution is from a reduced form of yttrium with a binding energy of 155.3 eV, and a second contribution is from an oxidized form of yttrium  at 157.6 eV that matches that of the air-exposed sample. A reduced form of yttrium is consistent with our understanding of Y$_2$CF$_2$, in which we expect the Y adopts a 2+ oxidation state. The second contribution is broadened compared to the air-exposed sample possibly due to the small amount of charging present during the collection of data. Therefore, we conclude that the surface of Y$_2$CF$_2$ oxidizes quickly during the brief transfer of the sample into the instrument and that the oxidation affects the reduced yttrium and carbidic carbon at the surface. Unfortunately, Ar-ion sputtering the surface seemed to damage the sample and only resulted in broadened peaks and more charging. 

\begin{figure}[h]
    \centering
    \includegraphics[width=16.5cm]{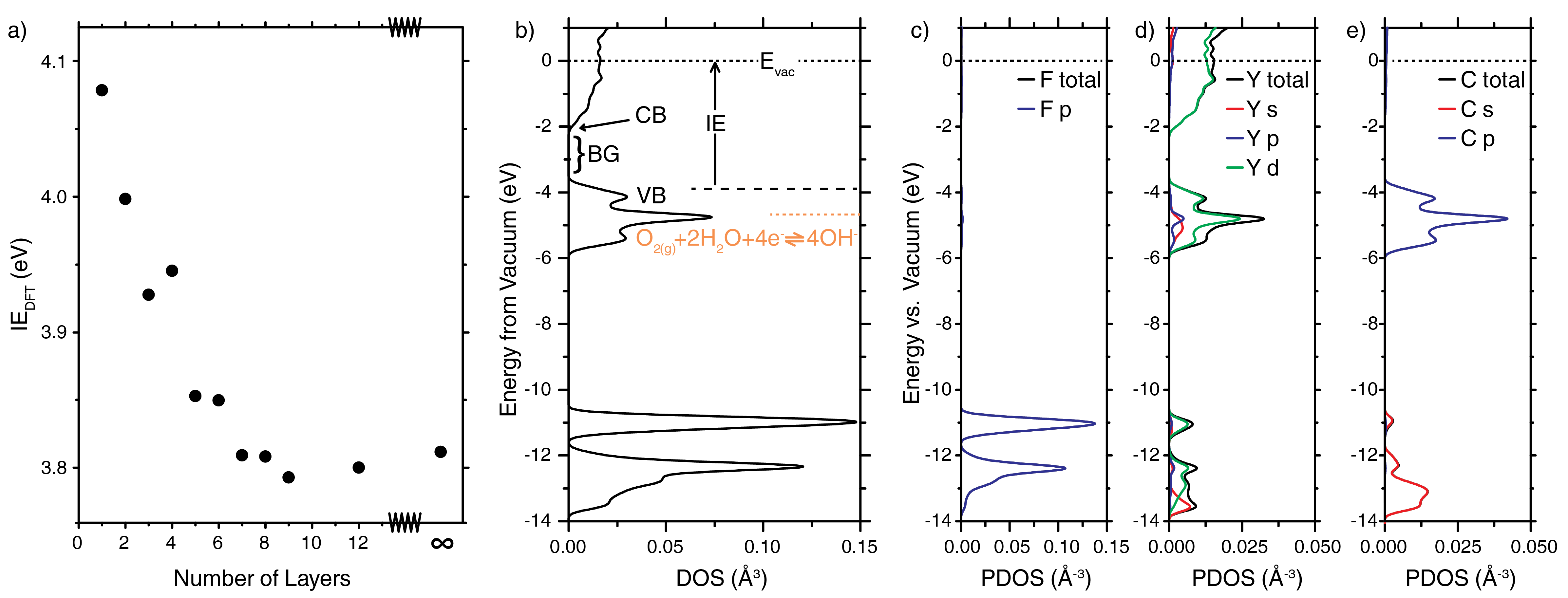}
    \caption[Band alignment of Y$_2$CF$_2$.]{ The band positions of Y$_2$CF$_2$ on an absolute scale. a) calculations of the ionization potential (IE) by DFT as described in the experimental for increasingly thick slabs of Y$_2$CF$_2$, which plateau at a value of $\sim$3.8 eV for a 12-slab and for a bulk crystal. b) The density of states (DOS) of bulk Y$_2$CF$_2$ calculated by DFT referenced to vacuum energy using the calculated ionization potential, showing the valence band (VB), conduction band (CB), and band gap (BG). The partial density of states for c) fluorine, d) yttrium, and e) carbon in Y$_2$CF$_2$.}
    \label{fig:Y2CF2_Bands}
\end{figure}

We return to the electronic structure to understand why Y$_2$CF$_2$ oxidizes in air. We place the band edges of Y$_2$CF$_2$ on an absolute scale along with the reduction potentials of oxygen and water, which are likely redox couples. We combined several measurements and calculations to place the band edges: DFT calculations gave the valence band edge (Figure \ref{fig:Y2CF2_Bands}a) and band gaps (Figure \ref{fig:Y2CF2_Optics}b,e), which are supported by our spectroscopic measurements (Figure \ref{fig:Y2CF2_Optics}a). The valence band edge is at a higher potential than the reduction potentials of oxygen and water, which explains why the material oxidizes. We gain further understanding of its reactivity through analysis of the partial density of states (PDOS), deconstructed into the elemental compositions and angular momentum (Figure \ref{fig:Y2CF2_Bands}c-e). The valence band consists of Y \textit{4d} and C \textit{2p} states. The C \textit{2p} states lie at high energies, and therefore can be readily donated. The diagram suggests electrons can be donated to oxygen, producing hydroxide.  If this initial oxidation is not self-limiting, more highly oxidized products may form, such as yttrium carbonate. This mechanism is consistent with the XPS observations of fast partial oxidation to an oxide or hydroxide after a brief exposure to air and a subsequent slow oxidation to a more oxidized form.

In general, MXenes have been considered stable in air and water; however, growing evidence reveals that MXenes decompose over the course of a few days to a few weeks.~\cite{Zhang_C2017, Li_G2019, Li_J2019, Huang_S2019, Chertopalov2018, Wang_X2017, Chae_Y2019, C8TA01468J} For example, Ti$_3$C$_2$ (-O, -OH, -F) decomposes to form TiO$_2$ in the presence of oxygen~\cite{Zhang_C2017, Chae_Y2019} and water.~\cite{Wang_X2017, Huang_S2019} In fact, the etching synthetic method, which uses strong oxidants like HF,~\cite{Naguib2011} H$_2$O$_2$,~\cite{ doi:10.1002/anie.201802232} or HCl/LiF,~\cite{Ghidiu2014} contributes to the oxidation, as evidenced by the formation of -O terminating moieties. The oxidation has consequences on the electronic structure.~\cite{ doi:10.1021/acs.jpcc.8b11525} For example, the work function of Ti$_3$C$_2$X (X = -O, -OH, -F) was predicted to vary from 1.9 to 6.2 eV depending on the surface terminations.~\cite{ Khazaei2015} Moreover, the work function of MXenes with random surface moieties depends on the complex interplay between the different surface moieties and their local dipoles.~\cite{Schultz2019} 

In this context, Y$_2$CF$_2$ and crystalline stacked MXenes provide an opportunity. The uniform, single F- moiety of crystalline stacked MXenes could facilitate experiments to understand and control its surface oxidation. Although we have observed that Y$_2$CF$_2$ oxidizes in air, the crystalline 3D MXene reported here is intrinsically oxygen-free, unlike the 2D MXenes reported so far.  It therefore provides a pristine material against which changes in surface composition and surface properties can be detected.  In addition, if this pristine 3D MXene can be exfolaited in water-free or oxygen-free conditions, it may also yield an unoxidized, pristine 2D MXene.

\section{Conclusion}

We have synthesized Y$_2$CF$_2$, a crystalline, 3D stack of MXene-like sheets capped exclusively by fluorine. Instead of the etching approach, we introduce a new synthetic method to directly synthesize the 3D crystal through a high-temperature solid-state reaction that ensures that only fluoride ions terminate the layers. Unlike MXenes, ionic bonds hold layers together. Interestingly, because the fluorine orbitals contribute to neither the valence band nor conduction band, the fluorine layers electrically isolate the yttrium carbide layers from each other. This imparts to the 3D crystal 2D-like electronic transport, and supports an understanding of these materials as a crystalline stack of 2D MXenes.The valence band edge, made of Y \textit{4d} and C \textit{2p} states, sits at a high potential, affording reducing character, which makes the crystal sensitive to air, but also suggests that the crystal could be useful as low work function electron emitters~\cite{Khazaei2017} or catalysts,~\cite{Seh_Z2016,Trasatti1972} in addition to applications in electronic,~\cite{Zhao_M2015, Li_R2017} sensing,~\cite{Kim_S2018,Yu_X2015} and energy storage.~\cite{Naguib2012,Liang_X2015,Wang_X2015}

\begin{acknowledgement}

S.C.W. acknowledges support of this research by NSF grant DMR-1905294. J.D.S and J.T.P. acknowledge support by the NSF Graduate Research Fellowship under grants DGE-1650114. This work was performed in part at the Chapel Hill Analytical and Nanofabrication Laboratory, CHANL, a member of the North Carolina Research Triangle Nanotechnology Network, RTNN, which is supported by the National Science Foundation, Grant ECCS-1542015, as part of the National Nanotechnology Coordinated Infrastructure, NNCI. The authors acknowledge the support of the UNC EFRC Center for Solar Fuels, an Energy Frontier Research Center funded by the U.S. Department of Energy, Office of Science, Office of Basic Energy Sciences under Award Number DE-SC0001011, for access to a Cary 5000 UV-Vis spectrometer with the DRA-2500 internal integrating sphere accessory instrumentation. The authors are also grateful to the Research Computing Center, University of North Carolina at Chapel Hill, for access to needed computing facilities to perform the computational studies reported in this work.

\end{acknowledgement}

\begin{suppinfo}

Additional details regarding the synthesis and characterization of Y$_2$CF$_2$, ball-milled Y$_2$CF$_2$, and air-exposed Y$_2$CF$_2$; SEM and EDS data, XRD, optical measurements of Y$_2$CF$_2$, photographs of powders. 

\end{suppinfo}


\providecommand{\latin}[1]{#1}
\makeatletter
\providecommand{\doi}
  {\begingroup\let\do\@makeother\dospecials
  \catcode`\{=1 \catcode`\}=2 \doi@aux}
\providecommand{\doi@aux}[1]{\endgroup\texttt{#1}}
\makeatother
\providecommand*\mcitethebibliography{\thebibliography}
\csname @ifundefined\endcsname{endmcitethebibliography}
  {\let\endmcitethebibliography\endthebibliography}{}








\end{document}